\documentstyle[12pt,fullpage]{article}
\def\stackunder#1#2{\mathrel{\mathop{#2}\limits_{#1}}}
\begin{document}
%  REMOVE THE '%' FROM THE NEXT 4 LINES TO INSERT A TAUP NUMBER
\begin{flushright}
TAUP-2669-2001
\end{flushright}
\vskip 3 true cm
\parskip=10pt
\baselineskip 7mm
\parindent=0cm 
\begin{center}
{\bf \Large The Covariant Stark Effect}

M. C. Land$^1$  and L. P. Horwitz$^2$

$^1$Department of Computer Science\\
Hadassah College\\
Jerusalem, Israel

$^2$School of Physics and Astronomy  \\
Raymond and Beverly Sackler Faculty of Exact Sciences  \\
Tel Aviv University, Ramat Aviv, Israel 

\end{center}
\vskip .5 true cm
\begin{abstract}
\parindent=0cm 
\parskip=10pt
This paper examines the Stark effect, as a first order
perturbation of manifestly covariant hydrogen-like bound states.
These bound states are solutions to a relativistic Schr\"odinger
equation with invariant evolution parameter, and represent mass
eigenstates whose eigenvalues correspond to the well-known energy
spectrum of the nonrelativistic theory.  In analogy to the
nonrelativistic case, the off-diagonal perturbation leads to a
lifting of the degeneracy in the mass spectrum.  In the covariant
case, not only do the spectral lines split, but they acquire an
imaginary part which is linear in the applied electric field,
thus revealing induced bound state decay in first order
perturbation theory.  This imaginary part results from the
coupling of the external field to the non-compact boost
generator.  In order to recover the conventional first order
Stark splitting, we must include a scalar potential term. This
term may be understood as a fifth gauge potential, which
compensates for dependence of gauge transformations on the
invariant evolution parameter.
\end{abstract}
\section{Introduction}
The Stark effect --- the splitting of degenerate spectral lines
in an electric field --- was an important early success for
quantum theory, and has remained a classroom staple, providing
the introduction to perturbation theory for degenerate states.
Paired with the Zeeman effect, in which an external magnetic
field couples to the diagonal (but otherwise degenerate) angular
momentum operator, the Stark effect demonstrates that this same
degeneracy rescues the first order perturbation from the coupling
of the external electric field to the off-diagonal position
operator.  Although the non-compact position operator cannot be
considered a small perturbation in any rigorous sense, and in
non-perturbative solutions, the discrete energy spectrum goes
over to a continuous resonance spectrum, \cite{resonance},
the first order nonrelativistic splitting is the basis for the
treatment of Stark broadening in spectroscopy.  Stark broadening
is been an important consideration in plasma physics
\cite{plasma} and has become a practical diagnostic tool in
surface science \cite{surface} and astronomy \cite{astronomy}.
The strong electric fields required to observe the effect
(Johannes Stark's 1913 observation was made with field strengths
of $10^5$ V/cm while typical fields may be two orders of
magnitude higher \cite{diagnostic}), suggest that a
relativistically covariant formulation of the problem may be
required, especially as the phenomenon is applied to high
precision measurement.

In this paper, we discuss the Stark effect as a first order
perturbation to a solution of the two body bound state problem in
relativistically covariant quantum mechanics.  This formulation
of the problem is based on Stueckelberg's off-shell kinematics
with invariant evolution parameter \cite{Stueckelberg},
generalized to the many particle case by Horwitz and Piron
\cite{H-P} (see also \cite{list}).  The relaxation of the
mass-shell constraint for particle kinematics is required to
achieve an action-at-a-distance framework with scalar potential.
In this framework, Arshansky and Horwitz \cite{I} obtained exact
solutions for relativistic generalizations of the classical
central force problems.  These wavefunctions form an induced
representation of the Lorentz group \cite{II}, and are degenerate
in the new quantum numbers associated with the enlarged symmetry.
Moreover, dipole radiation, emitted in transitions among these
bound states, obeys selection rules which are formally identical
to those of the nonrelativistic problem but with covariant
interpretation \cite{selrul}.  The bound state solutions for the
Coulomb problem represent mass eigenstates whose eigenvalues
correspond to the well-known energy spectrum of the
nonrelativistic theory \cite{I}.

The covariant Zeeman effect has been previously obtained
\cite{zeeman} and the covariance of the approach permits
the application of machinery developed there to the Stark effect.
The construction of the action for the induced representation
requires care, especially the coupling to the vector
field in a manner which preserves both Lorentz and local gauge
invariance.  In the case of
constant external electromagnetic field, the first order
interaction term becomes a scalar contraction of the field
strength tensor with the Lorentz generators.  The Zeeman effect
is then recovered as a magnetic-like field coupled to the rotation
generators, and the Stark effect is obtained as an electric-like field
coupled to the boost generators.  Since the non-compact boost
generators have complex eigenvalues, the relativistic
bound states decay even at first order.  To recover
the usual Stark splitting, we must include an external scalar
potential involving a coupling to the spacetime position
four-vector.  This `fifth potential' has a natural interpretation
in the pre-Maxwell electromagnetic theory \cite{saad}, where it
plays the role of a gauge field compensating for transformations
which depend on the invariant evolution parameter.  In the
pre-Maxwell theory, the photon kinematics are also off-shell,
however the measurement process picks out the zero-mass
eigenstate as an equilibrium state \cite{concat}.  Under this
interpretation of the Stark effect, the off-shell photon becomes
a necessary corollary to the parameterized quantum mechanics
formalism.

The Stueckelberg equation for the two body problem,
\begin{equation}
i\partial_\tau \psi(x_1,x_2,\tau) = K \psi(x_1,x_2,\tau) =
\left[\frac{p_{1\mu}p_1^{\mu}}{2M_1} + \frac{p_{2\mu}p_2^{\mu}}{2M_2} 
+ V(x_1,x_2) \right] \psi(x_1,x_2,\tau)
\label{two_body}
\end{equation}
is Poincar\'e invariant and quadratic in the four momenta.  
The nonrelativistic central force problems may be generalized to
covariant form \cite{I} through the replacement
\begin{equation}
r = \sqrt{({\bf r}_1 - {\bf r}_2)^2} \qquad \longrightarrow \qquad
\rho = \sqrt{({\bf r}_1 - {\bf r}_2)^2 - (t_1 -t_2)^2}
\label{replacement}
\end{equation}
in the argument of the usual potentials.  Since $t_1
\rightarrow t_2 $ in the Galilean limit, the original 
nonrelativistic problem is recovered in this limit.

One may separate variables of the center of mass
motion and relative motion in the same way as in the
nonrelativistic theory,
\begin{equation}
K = {P^{\mu} P_{\mu} \over 2M} + {p^{\mu} p_{\mu}
\over 2m} + V(\rho),
\label{separate}
\end{equation}
where
\begin{equation}
P^\mu = p_1^\mu + p_2^\mu \qquad\qquad M = M_1 + M_2
\label{c_m_def}
\end{equation}
$$p^\mu = (M_2p_1^\mu - M_1 p_2^\mu)/M \qquad\qquad m=M_1M_2/M.$$
The reduced motion is then described by the relative Hamiltonian 
\begin{equation}
K_{rel} = {p^{\mu} p_{\mu} \over 2m} + V(\rho) \ \ .
\label{relative}
\end{equation}

In order to obtain the correct nonrelativistic limit for the
spectrum in the Coulomb problem, one must
choose an arbitrary spacelike unit vector $n_\mu$ ($g_{\mu\nu} = {\rm
diag}(-1,1,1,1)\; \Rightarrow \; n^2=+1$) and restrict the
spacetime support of the eigenfunctions to a Restricted Minkowski Space
(RMS) corresponding to the condition
\begin{equation}
(x_\perp )^2 = [x-(x \cdot n) n ]^2 \geq 0,
\label{RMS}
\end{equation}
where $x \equiv x^{\mu}$ is the {\it relative}
coordinate $x^\mu_1 - x^\mu_2$,
and $x^2 = x^\mu x_\mu$.  The RMS is transitive and invariant under
the O(2,1) subgroup of O(3,1) leaving $n_\mu$ invariant
and translations along $n_\mu$.  The choice of $n_\mu$ along the $z$-axis
leads to the parameterization 
$$y^0 = \rho \; \sinh \beta \; \sin \theta
\qquad\qquad y^1 = \rho \; \cosh \beta \; \sin \theta \;
\cos \phi$$
\begin{equation}
y^2 = \rho \; \cosh \beta \; \sin \theta \; \sin \phi
\qquad\qquad y^3 = \rho \; \cos \theta 
\label{RMS_coordinates}
\end{equation}
for which 
\begin{equation}
(y^1)^2 + (y^2)^2 - (y^0)^2 \geq 0.
\label{RMS_rule}
\end{equation}
The eigenfunctions of $K_{rel}$
form irreducible representations of SU(1,1) --- in the double
covering of O(2,1) --- parameterized by the spacelike vector
$n_\mu$ stabilized by the particular O(2,1) \cite{I,II}.  

An induced representation of SL(2,C) was constructed \cite{II}, by applying
the Lorentz group to the RMS coordinates $x^\mu$ and the frame
orientation $n_\mu$, and studying the action on these
wavefunctions.  A set of wavefunctions with support on $(n,x)$
where
\begin{equation}
x \in {\rm RMS}(n_\mu) = \left\{ x \ | \ [x-(x \cdot n) n
]^2 \geq 0 \right\}
\label{support}
\end{equation}
may be regarded as functions of the chosen $n_\mu$ and the coordinates
of a standard frame $y \in {\rm RMS}({\mathaccent'27n}_\mu)$,
since the Lorentz transformation ${\cal L}$ which performs the
mapping $ {\mathaccent'27n} = {\cal L} (n) \; n $ has the
property that
\begin{equation}
x \in {\rm RMS}(n_\mu) \qquad {\rm and} \qquad y = {\cal L} (n) \;
x \qquad \Longrightarrow \qquad y \in {\rm RMS}
({\mathaccent'27n}_\mu).
\label{rule}
\end{equation}
For the choice ${\mathaccent'27n} = (0,0,0,1)$, the
parameterization (\ref{RMS_coordinates}) may be used for $y^\mu$,
and the effect on the wavefunctions of a Lorentz
transformation $\Lambda$, may be seen from the composition
\begin{equation}
\begin{array}{ccc}
x\in {\rm RMS}\left( n_{\mu }\right)  &
\stackrel{\Lambda }{\longrightarrow } &
x^{\prime }\in {\rm RMS}\left( n_{\mu }^{\prime }\right)  \\ 
&  &  \\ 
\uparrow {\cal L}(n)^{T} &  &
\downarrow {\cal L}(\Lambda n) \\ 
&  &  \\ 
y\in {\rm RMS} \left( \stackrel{\circ }{n}_{\mu }\right)  &  &
y^{\prime }\in {\rm RMS} \left( \stackrel{\circ }{n_{\mu }}\right) 
\end{array}
\end{equation}
to be
\begin{equation}
\psi_n (y) \rightarrow \psi_n^\Lambda (y) = \psi_{\Lambda^{-1}n}
(D^{-1} (\Lambda^{-1} , n) \; y) 
\label{D-matrix}
\end{equation}
where $\Lambda$ acts directly on $n_\mu$.  The representations  
are moved on an orbit generated by this spacelike vector, and  
the Lorentz transformations act on $y^\mu$ through the O(2,1)  
little group, represented by $D^{-1} (\Lambda , n)$, with the  
property 
\begin{equation}
D^{-1} (\Lambda , n) \; {\mathaccent'27n} = {\cal L} (\Lambda n) \;
\Lambda \; {\cal L}^T (n) \; {\mathaccent'27n} \equiv
{\mathaccent'27n} .
\label{little_group}
\end{equation}
Expressing the matrix Lorentz generators as
\begin{equation}
({\cal M}^{\sigma\lambda})^{\mu\nu} = g^{\sigma\mu}g^{\lambda\nu} -
g^{\sigma\nu}g^{\lambda\mu},
\label{eqn:1.14}
\end{equation}
the matrix ${\cal L} ^T(n)$ was chosen in \cite{II} to be 
\begin{eqnarray}
{\cal L} ^T(n)  &=& e^{\gamma {\cal M}^{23}}e^{\omega
{\cal M}^{31}}e^{\alpha {\cal M}^{03}}
\label{eqn:1.13}
\\
\nonumber\\
&=& \pmatrix{\cosh \alpha & 0 & 0 & \sinh \alpha
\cr - \sin\omega\; \sinh\alpha & \cos\omega & 0 & - \sin\omega\;  
\cosh\alpha \cr \sin\gamma \; \cos\omega \; \sinh\alpha &
\sin\gamma \; \sin\omega & \cos\gamma & \sin\gamma \; \cos\omega
\; \cosh\alpha \cr \cos\gamma \; \cos\omega \; \sinh\alpha &  
\cos\gamma \; \sin\omega & -\sin\gamma & \cos\gamma \;  
\cos\omega \; \cosh\alpha \cr } , 
\label{eqn:1.15}
\end{eqnarray}
which provides the parameterization of $n_\mu$ as
\begin{equation}
n_\mu = \pmatrix{\sinh \alpha \cr -\sin\omega\; \cosh\alpha \cr 
\sin\gamma \; \cos\omega \; \cosh\alpha \cr \cos\gamma \;
\cos\omega \; \cosh\alpha \cr } .
\label{eqn:1.16}
\end{equation}
The generators $h_{\alpha\beta}(n)$ of
(\ref{D-matrix}) form a representation of the O(3,1) Lie
algebra (through their action on $y$ and $n$), and the Casimir
operators
\begin{equation}
\hat{c}_1= {1\over 2} h_{\alpha\beta}(n) h^{\alpha\beta}(n)
\qquad \hat{c}_2 =
{1\over 2} \epsilon^{\alpha\beta\gamma\delta} h_{\alpha\beta}(n)
h_{\gamma\delta}(n) 
\label{eqn:1.17}
\end{equation}
and the operators of the SU(2) subgroup 
\begin{equation}
{\bf L}^2 (n) = {1\over 2} h_{ij}(n)h^{ij}(n) \qquad L_1 (n) 
= h^{23} (n) = -i {\partial \over \partial \gamma} 
\label{eqn:1.18}
\end{equation}
can be constructed as a commuting set.  Moreover, the operator 
\begin{equation}
\Lambda = {1 \over 2}M^{\mu \nu} M_{\mu \nu} \rightarrow
\ell(\ell +1) - {3\over 4},
\label{eqn:1.19}
\end{equation}
where $M^{\mu \nu} = y^\mu p^\nu - y^\nu p^\mu$, and the O(2,1)
Casimir $N^2 = (M^{01})^2 + (M^{02})^2 + (M^{12})^2 $
commute with this set.  The wavefunctions which are
eigenfunctions of the set
\begin{equation}
\{\Lambda , N^2, \hat{c}_1, \hat{c}_2 , {\bf L}^2 (n) , L_1 (n) \}
\label{eqn:1.20}
\end{equation}
with eigenvalues $Q= \{ \ell(\ell +1) - {3 \over 4}, n^2  
-{1\over 4 }, c_1, c_2, L(L+1), q \}$ form a representation of
SL(2,C).  The requirement that  
the resulting representation be unitary and irreducible  
(the wavefunctions lie in the principal series), imposes the  
condition $c_1 = \hat{n} ^2 -1 - c_2^2 / \hat{n} ^2 $, where  
$\hat{n} = n+1/2$.  

The wavefunctions in the induced representation have the explicit
form \cite{I}
\begin{equation}
 \psi_n^Q (y) = R_{n_a \ell}(\rho) \; \Theta_{\ell}^n(\theta) \;
\xi^Q (n_\mu , \beta , \phi ) 
\end{equation}
where
\begin{equation}
\Theta_{\ell}^n(\theta) = (1-\xi^2)^{- {1\over 4}} \sqrt{{2{\ell}+1\over 2}
{({\ell}-n)! \over ({\ell}+n)!} } \; P_{\ell}^n (\xi ) 
\end{equation}
\begin{equation}
 \xi^Q (n_\mu , \beta , \phi ) = \sum_{k=0}^{L-\hat{n}}
{\cal D}_k^Q (\alpha ,
\omega , \gamma ) \; \chi_{n+k}^{-n} (\beta , \phi) 
\end{equation}
\begin{equation}
\chi_{n+k}^{-n} (\beta , \phi) = B_{n+k,n} (\beta) \; \Phi_{n+k}
(\phi) 
\end{equation}
\begin{equation}
B_{n+k,n} (\beta) = (1-\zeta^2)^{1\over 4} \sqrt{ n {(2n+k)!\over
k! }} \; P_{n+k}^{-n} (\zeta) 
\end{equation}
\begin{equation}
\Phi_{n+k} (\phi) = {1\over \sqrt{2\pi} } e^{i(n+k+ {1\over 2})
\phi} 
\end{equation}
\begin{equation}
{\cal D}_k^Q (\alpha ,\omega , \gamma ) = \Xi_{Lk}^{nc_2} (u)
\; P_{q,-M_k}^{L} (z) e^{-iq\gamma} 
\end{equation}
\begin{equation}
\Xi_{Lk}^{nc_2} (u) =(-1)^k \sqrt{ (2\hat{n} +k-1)! \over (2\hat
n -1)!k!} N_L^Q (1-u^2)^{-{\hat{n} -1 \over 2}} \; P_{-{ic_2 \over
\hat{n}},\hat{n} +k}^L (u) 
\end{equation}
with $u=\tanh \alpha $, $z=\sin \omega$, $\xi=\cos \theta $,
$\zeta=\tanh \beta $, $M_k = \hat{n} +k$ and $N_L^Q$ a normalization
constant.  The functions $P_{\ell}^n (\xi )$ are standard
Legendre polynomials, and $P^L_{ab}$ is related to the Jacobi
polynomials $P_k^{\alpha\beta}$ through
\begin{equation}
P^L_{ab}(z) = \frac{i^{a-b}}{2^a} 
\sqrt{\frac{(L-a)!(L+a)!}{(L-b)!(L+b)!} }
(1-z)^{\frac{a-b}{2}}(1+z)^{\frac{a+b}{2}}
P_{L-a}^{(a-b,a+b)}(z)
\label{}
\end{equation}

These wavefunctions are orthogonal with respect to the measure 
$d^4 y \; d^4 n\; \delta (1-n^2)$, where
\begin{eqnarray}
\int d^4 y &=& \int_{0}^{\infty} d\rho \; \rho^3
\int_{-\infty}^{\infty} d\beta \; \cosh \beta \int_{0}^{\pi}
d\theta \; \sin ^2 \theta \int_{0}^{2\pi} d\phi
\nonumber \\
&=& \int_{0}^{\infty} d\rho \; \rho^3 \int_{-1}^{1} d\xi \;
\sqrt{1-\xi^2} \int_{-1}^{1} d\zeta \; (1-\zeta^2)^{-{3\over
2}} \int_{0}^{2\pi} d\phi
\label{orth1}
\end{eqnarray}
\begin{eqnarray}
\int d^4 n \; \delta (1-n^2) &=& {1\over 2} \int_{-\infty}^{\infty}
d\alpha \cosh ^2 \alpha \int_{-{\pi \over 2}}^{\pi \over 2}
d\omega \cos \omega \int_{0}^{2\pi} d\gamma
\nonumber \\
&=& {1\over 2} \int_{-1}^{1} {du\over (1-u^2)^2}
\int_{-1}^{1} dz \int_{0}^{2\pi} d\gamma 
\label{orth2}
\end{eqnarray}
The remaining ``radial'' function, after the transformation  
$\hat R (\rho) = \sqrt{\rho} R (\rho) $ must satisfy an
equation which is precisely of  
the form of the nonrelativistic Schr\"odinger radial equation in  
three dimensions (and has the same normalization).  The states  
$\psi_n (y)$ are then eigenstates of the Lorentz invariant  
$K_{rel}$, whose support is on the RMS($n$), with the quantum  
numbers (\ref{eqn:1.20}), and a principal quantum number $n_a$.   
In particular, the solutions for the problem corresponding to  
the Coulomb potential \cite{I} yield bound states with a mass  
spectrum which coincides with the nonrelativistic Schr\"odinger  
energy spectrum.  

\section{Phase Space}
The Coulomb interaction has support in the RMS of an arbitrary
unit vector $n_\mu$.  However, it was shown in \cite{selrul} that
under dipole emission, the shift in the eigenvalue of $L_1 (n)$
corresponds to a recoil in the orientation of $n_\mu$ with
respect to the polarization of the emitted or absorbed photon.
The dependence of the magnetic quantum number $q$ on the frame
orientation is not surprising, since the operator $L_1
(n)$ belongs to the SU(2) subgroup of SL(2,C), and acts on
$n_\mu$, but not on the RMS coordinates (it was shown in
\cite{selrul} that for $\Lambda$ a rotation about the 1-axis,
${D^{-1} (\Lambda , n)} \equiv 1$).

In order to consider the coupling to an external electromagnetic
field, we construct a classical Lagrangian, in which $n_\mu$
plays an explicit dynamical role along with the RMS coordinates
$x_\mu$.  We show that the Lorentz generators are conserved
quantities for this action, and construct the Hamiltonian, which
may be unambiguously quantized and made locally gauge invariant.

We first consider the classical phase space parameterized by
$(n,y)$ and their $\tau$-derivatives.  From the known
transformation properties,
\begin{equation}
n \rightarrow n' = \Lambda \; n \qquad x \rightarrow x' =
\Lambda \; x 
\label{eqn:2.1}
\end{equation}
we find that
\begin{equation} 
x' = \Lambda \; x = \Lambda \ \left({\cal L} (n)^T \; y\right) = 
\left({\cal L} (\Lambda n)^T {\cal L} (\Lambda n)\right)\Lambda \; 
{\cal L} (n)^T \; y = {\cal L} (n')^T \; y' .
\label{eqn:2.2}
\end{equation}
so that $y$ transforms as 
\begin{equation}
y \rightarrow y' = {D^{-1} (\Lambda , n)} \; y ,
\label{eqn:2.3}
\end{equation}
where ${D^{-1} (\Lambda , n)}={\cal L}
(\Lambda n)\; \Lambda \;{\cal L}(n)^{T} $ belongs to the O(2,1)
which leaves ${\mathaccent'27n}$ invariant,
i.e.,
\begin{equation}
D^{-1} (\Lambda , n) \; {\mathaccent'27n} = {\cal L} (\Lambda
n)\; \Lambda \; {\cal L} (n)^T \; {\mathaccent'27n} =
{\mathaccent'27n} \ .
\label{eqn:2.4}
\end{equation}
The coordinates thus transform as 
\begin{equation}
\Lambda :  \; (n,y) \quad \rightarrow \quad (n,y)' = (\Lambda  n ,
D^{-1} (\Lambda , n) y).
\label{eqn:2.5}
\end{equation}
Since $\tau$ is a scalar invariant, the {\it velocity} $\dot n = dn/d\tau$ 
transforms as a vector,
\begin{equation}
n' =\Lambda \; n \quad \Longrightarrow \quad \dot n' =
\Lambda \; \dot n 
\label{eqn:2.7} \ .
\end{equation}
However ${\cal L} (n)$ is now $\tau$-dependent through $n_\mu$, so that
\begin{eqnarray}
y= {\cal L} (n(\tau)) \; x \quad &\Longrightarrow& \quad \dot y =
{\cal L} (n) \dot x + \dot {\cal L} (n) x \label{eqn:2.8}\\
x= {\cal L} (n(\tau))^T \; y \quad &\Longrightarrow& \quad
\dot x = {\cal L} (n)^T \dot y + \dot {\cal L} (n)^T y \ .
\label{eqn:2.9}
\end{eqnarray}
But since $d \Lambda /d\tau =0$, (\ref{eqn:2.8}) is nevertheless form
invariant:
\begin{eqnarray}
(\dot y)' &=& {\cal L} (n') \dot x' + \dot {\cal L} (n') x'
\nonumber \\
&=& {\cal L} (\Lambda n) [\Lambda \dot x ] + \dot {\cal L}
(\Lambda n)[\Lambda x]
\nonumber \\
&=& {\cal L} (\Lambda \; n) \Lambda [{\cal L} (n)^T \dot 
y + \dot {\cal L} (n)^T y ] + \dot {\cal L} (\Lambda n)
[\Lambda {\cal L}(n))^T \; y ]
\nonumber \\
&=& [ {\cal L} (\Lambda n) \Lambda {\cal L} (n)^T ] \dot y + 
[ {\cal L} (\Lambda n) \Lambda \dot {\cal L} (n)^T  + \dot
{\cal L} (\Lambda \; n) \Lambda {\cal L} (n))^T] \; y  
\nonumber \\
&=& D^{-1} (\Lambda , n) \dot y + \dot D^{-1} (\Lambda , n) \; y
\nonumber \\
&=& \frac{d}{d\tau} [ D^{-1} (\Lambda , n) \; y ] .
\label{eqn:2.10}
\end{eqnarray}
In summary, the {\it phase space} transforms as:
\begin{equation}
\Lambda: \quad \{ (n,y);(\dot n,\dot y)\} \longrightarrow 
\{ (\Lambda n, D^{-1} (\Lambda , n) y ); (\Lambda \dot n, 
D^{-1} (\Lambda , n) \dot y+ \dot D^{-1} (\Lambda , n) y) \} \ .
\label{eqn:2.11}
\end{equation}

To obtain the classical generators of the Lorentz
transformation (\ref{eqn:2.5}), we expand the
matrix form of the Lorentz transformations as 
\begin{equation}
\Lambda =1+ \lambda +o(\lambda ^2)
\label{eqn:2.12}
\end{equation}
and write $\lambda$ as
\begin{equation}
\lambda = \frac{1}{2} \; \omega_{\alpha\beta} \;
{\cal M}^{\alpha\beta}
\label{eqn:2.13}
\end{equation}
where $\omega_{\alpha\beta}, \; \alpha,\beta =0, \cdots,3$
is (infinitesimal) antisymmetric.  The matrix generators 
\begin{equation}
{\cal M}^{\alpha\beta} =  \left.  \frac{\partial \lambda}{\partial
\omega_{\alpha\beta}} \right|_{\omega=0} 
\label{eqn:2.14}
\end{equation}
are those given in (\ref{eqn:1.14}).
According to (\ref{eqn:2.12}) and (\ref{eqn:2.13}),
(\ref{eqn:2.5}) becomes 
\begin{equation}
\Lambda :  \; (n,y) \quad \rightarrow \quad (n,y)' = (n+ \lambda 
n , {\cal L}(n+\lambda n )(1+\lambda ) {\cal L}(n)^T y) +
o(\omega ^2).
\label{eqn:2.15}
\end{equation}
Representing the classical generators of $\xi=(n,y) \rightarrow
\xi'=(n',y')$ as
\begin{equation}
X_{\alpha\beta} = \sum_{i=1}^{8} \left. \frac{\partial
\xi^i}{\partial \omega^{\alpha\beta}} \right|_{\omega=0}
\frac{\partial}{\partial \xi^i}
\label{eqn:2.16}
\end{equation}
where
\begin{equation}
\xi ^{i}=\left\{ 
\begin{array}{c}
n^{\mu }\quad {\rm  for }\quad i=1,\cdots ,4, \quad \mu=0,\cdots,3 \\ 
\\ 
y^{\mu }\quad {\rm  for }\quad i=5,\cdots ,8, \quad \mu=0,\cdots,3 
\end{array}
\right. 
\end{equation}
we obtain for $i=1,\cdots,4$,
\begin{equation}
\sum_{i=1}^{4} \left. \frac{\partial \xi^i}{\partial
\omega^{\alpha\beta}} \right|_{\omega=0}
=({\cal M}_{\alpha\beta})^\mu_{\ \nu} n^\nu
\frac{\partial}{\partial n^\mu}
=n_\beta \frac{\partial}{\partial n^\alpha} -
n_\alpha \frac{\partial}{\partial n^\beta}
\label{eqn:2.18}
\end{equation}
which was called $d(\lambda_{\alpha\beta})$ in \cite{II}. 
Similarly, for $i=5,\cdots,8$,
\begin{eqnarray}
\left. \frac{\partial \xi^i}{\partial \omega^{\alpha\beta}} 
\right|_{\omega=0} &=& \left. \frac{\partial}{\partial 
\omega^{\alpha\beta}} \left[{\cal L} (n+\lambda n) (1+\lambda)
{\cal L} (n)^T y \right] ^i \right|_{\omega=0}
\nonumber \\
&=& {\cal L} _{\sigma\beta} {\cal L} ^{\rho}_{\ \alpha}
(y^\sigma \frac{\partial}{\partial y^\rho} - y^\rho
\frac{\partial}{\partial y^\sigma} ) - n_\beta
{\cal L}^\rho_{\ \zeta}
\frac{\partial}{\partial n^\alpha} {\cal L}_\sigma^{\ \zeta}
(y^\sigma 
\frac{\partial}{\partial y^\rho} - y^\rho
\frac{\partial}{\partial y^\sigma} )
%
%\label{eqn:2.19}
\label{eqn:2.22}
\end{eqnarray}
which was called $g(\lambda_{\alpha\beta})$ in \cite{II}.
We have used the fact that 
\begin{equation}
{\cal L} (n) {\cal L} (n)^T =1 \quad \Longrightarrow \quad
\left(\frac{\partial}
{\partial n^\mu} {\cal L} (n) \right) {\cal L} (n)^T  +
{\cal L} (n) 
\frac{\partial}{\partial n^\mu} {\cal L} (n) ^T =0 .
\label{eqn:2.20}
\end{equation}
Finally, we obtain for the classical generators
\begin{equation}
X_{\alpha\beta} =  {\cal L} _{\sigma\beta}
{\cal L}^{\rho}_{\ \alpha} (y^\sigma 
\frac{\partial}{\partial y^\rho} - y^\rho
\frac{\partial}{\partial y^\sigma} ) - n_\beta
{\cal L} ^\rho_{\ \zeta}
\frac{\partial}{\partial n^\alpha} {\cal L}_\sigma^{\ \zeta}
(y^\sigma 
\frac{\partial}{\partial y^\rho} - y^\rho
\frac{\partial}{\partial y^\sigma} )  + n_\beta
\frac{\partial}{\partial 
n^\alpha} - n_\alpha \frac{\partial}{\partial n^\beta} 
\label{eqn:2.23}
\end{equation}
which was called $ih_n (\lambda_{\alpha\beta})$ in \cite{II},
and shown to satisfy the Lie algebra
of SL(2,C).  It is useful to maintain the matrix notation for
${\cal M}_{\alpha\beta}$ so that (\ref{eqn:2.23}) may be
written as
\begin{eqnarray}
X_{\alpha\beta} &=& [ {\cal L} (n) {\cal M}_{\alpha\beta}
{\cal L} ^T ]^\mu_{\ \nu} y^\nu 
\frac{\partial}{\partial y^\mu} -
[{\cal L} ({\cal M}_{\alpha\beta})^\rho_{\ \sigma} n^\sigma
\frac{\partial}{\partial n^\rho} {\cal L} ^T]^\mu_{\ \nu} y^\nu 
\frac{\partial}{\partial y^\mu} -
({\cal M}_{\alpha\beta})^\rho_{\ \sigma} n^\sigma
\frac{\partial}{\partial n^\rho} \nonumber \\
&=& - y^T [ {\cal L} (n) {\cal M}_{\alpha\beta} {\cal L} ^T ] 
\nabla_{\bf y} - y^T {\cal L} (n) [n^T {\cal M}_{\alpha\beta} 
\nabla_{\bf n} ] {\cal L} ^T \nabla_{\bf y} - n^T
{\cal M}_{\alpha\beta} \nabla_{\bf n} 
\label{eqn:2.24}
\end{eqnarray}
where $(\nabla_{\bf y})_\mu = \frac{\partial}{\partial y^\mu}$. 
By defining the four matrices
\begin{equation}
S_\mu = {\cal L} \frac{\partial}{\partial n^\mu}
{\cal L} ^T \qquad\qquad \mu = 0,\cdots,3
\label{eqn:2.25}
\end{equation}
(which by (\ref{eqn:2.20}) are antisymmetric) equation
(\ref{eqn:2.24}) becomes
\begin{equation}
X_{\alpha\beta} = -\left\{ y^T [ {\cal L} (n)
{\cal M}_{\alpha\beta} {\cal L}^T ] 
\nabla_{\bf y} + n_\mu ({\cal M}_{\alpha\beta})^{\mu\nu}
[ y^T S_\nu \nabla_{\bf y}
+ (\nabla_n)_\nu ] \right\}
\label{eqn:2.26}
\end{equation}
In the matrix notation of (\ref{eqn:2.26}), the generators found
in \cite{II} have the form
\begin{equation}
d_{n}(\lambda )=-n_{\mu }({\cal M}_{\alpha \beta })^{\mu \nu }(\nabla
_{n})_{\nu }
\end{equation}
\begin{equation}
g_{n}(\lambda )=-\left\{ y^{T}[{\cal L}(n){\cal M}_{\alpha \beta }{\cal L}
^{T}]\nabla _{{\bf y}}+n_{\mu }({\cal M}_{\alpha \beta })^{\mu \nu
}y^{T}S_{\nu }\nabla _{{\bf y}}\right\}
\end{equation}

For the action in $(n,y)$ coordinates, we choose the simplest
Lagrangian containing a kinetic term for $n_\mu$, which is
\begin{equation}
{\rm L}=\frac{1}{2}m\dot{x}^{2}+\frac{1}{2}mr_{0}^{2}\dot{n}^{2}-V(n,x) \ ,
\label{eqn:3.1}
\end{equation}
where the scale factor $r_{0}$ is required because $n_\mu$ is a
unit vector.  Using (\ref{eqn:2.9}) to expand $\dot{x}$,
we may write (\ref{eqn:3.1}) in the form
\begin{equation}
{\rm L}=\frac{1}{2}m[\dot{y}+{\cal L}\dot{{\cal L}}^{T}y]^{2}+\frac{1}{2}
mr_{0}^{2}\dot{n}^{2}-V(n,{{\cal L}}^{T}y) \ .  
\label{eqn:3.3}
\end{equation}
Notice that when $\dot n =0$, the dynamics depend only on
$\dot y$ and so the relative coordinate
remains within RMS($n$).  
By construction, (\ref{eqn:3.3}) is Lorentz invariant, and so
is invariant under the transformations induced by
(\ref{eqn:2.26}).  Therefore, applying Noether's theorem and the
Euler-Lagrange equation,
\begin{equation}
0=\delta {\rm L} = \frac{\partial{\rm L} }{\partial \xi ^i}
\delta \xi^i 
+ \frac{\partial{\rm L} }{\partial \dot \xi ^i}
\delta \dot \xi^i
= \left[ \frac{\partial{\rm L} }{\partial \xi ^i}
- \frac{d}{d \tau} 
\frac{\partial{\rm L} }{\partial \dot \xi ^i} \right]
\delta \xi^i + \frac{d}{d \tau}
\left[ \frac{\partial{\rm L}}{\partial \dot \xi ^i} 
\delta \xi^i \right],
\label{eqn:3.4}
\end{equation}
for the variation 
$\delta \xi^i = \frac{1}{2} \omega^{\alpha\beta}
X_{\alpha\beta} \ \xi^i$, one obtains the conservation law
\begin{equation}
\frac{d}{d \tau} [ {\rm p}^\mu X_{\alpha\beta} y_\mu +
\pi^\mu X_{\alpha\beta} n_\mu ] =0
\label{eqn:3.5}
\end{equation}
where 
\begin{equation}
{\rm p}_\mu = \frac{\partial{\rm L} }{\partial \dot y^\mu}
\qquad {\rm and}
\qquad \pi_\mu = \frac{\partial{\rm L} }{\partial \dot n^\mu} .
\label{eqn:3.6}
\end{equation}
Using
(\ref{eqn:2.26}) for $X_{\alpha\beta}$, (\ref{eqn:3.5}) becomes,
\begin{equation}
\frac{d}{d \tau} \{ y^T {\cal L} (n) {\cal M}_{\alpha\beta}
{\cal L} ^T {\rm p} + n_\mu
({\cal M}_{\alpha\beta})^{\mu\nu}
[ y^T S_\nu {\rm p} + \pi_\nu ] \}  =0.
\label{eqn:3.7}
\end{equation}
If we understand $\pi_\nu$, in the Poisson bracket sense,
as a derivative with respect to $n_\mu$, then
the quantum operators $h_n(\lambda_{\alpha\beta})$ of \cite{II}
now appear as classical constants of the motion
for the Lagrangian (\ref{eqn:3.1}).

To obtain the Hamiltonian, we first observe
that ${\cal L}$ depends on $\tau$ only through $n$, so
\begin{equation}
{\cal L} \dot {\cal L}^T = {\cal L} \left(\dot n^\nu
\frac{\partial}{\partial n^\nu}
{\cal L}^T\right) = \dot n^\nu S_\nu
\label{eqn:3.8}
\end{equation}
Applying (\ref{eqn:3.6}) to (\ref{eqn:3.3}), 
\begin{equation}
{\rm p}_\mu = \frac{\partial{\rm L} }{\partial \dot y^\mu} =
m[\dot
y_\mu + ({\cal L} \dot {\cal L}^T y)_\mu] \quad \Rightarrow
\quad {\rm p}=m[\dot y + \dot n^\nu S_\nu y]
\label{eqn:3.9}
\end{equation}
and
\begin{equation}
\pi _{\mu }=\frac{\partial {\rm L}}{\partial \dot{n}^{\mu }}=mr_{0}^{2}\dot{n
}_{\mu }+m[\dot{y}+\dot{n}^{\nu }S_{\nu }y]^{T}\frac{\partial }{\partial 
\dot{n}^{\mu }}[\dot{y}+\dot{n}^{\nu }S_{\nu }y]=mr_{0}^{2}\dot{n}_{\mu
}-y^{T}S_{\mu }{\rm p}  
\label{eqn:3.10}
\end{equation}
where we used (\ref{eqn:3.9}) and the antisymmetry of $S_\mu$
to obtain (\ref{eqn:3.10}). 
Equations (\ref{eqn:3.9}) and (\ref{eqn:3.10}) may be inverted
to eliminate $(\dot n , \dot y)$:
\begin{equation}
\dot{n}_{\mu }=\frac{1}{mr_{0}^{2}}[\pi _{\mu }+y^{T}S_{\mu }{\rm p}]
\label{eqn:3.11}
\end{equation}
and
\begin{equation}
\dot y = \frac{1}{m} {\rm p} - \dot n^\mu S_\mu y = \frac{1}{m}
\dot{y}=\frac{1}{m}{\rm p}-\dot{n}^{\mu }S_{\mu }y=\frac{1}{m}{\rm p}-\frac{1
}{mr_{0}^{2}}[\pi ^{\mu }+y^{T}S^{\mu }{\rm p}]S_{\mu }y  
\label{eqn:3.12}
\end{equation}
which may be used to write the Hamiltonian as
\begin{eqnarray}
{\rm K} &=&\dot{y}\cdot {\rm p}+\dot{n}\cdot \pi -{\rm L}  \nonumber \\
&=&\frac{{\rm p}^{2}}{2m}+\frac{1}{2mr_{0}^{2}}(\pi ^{\mu }+y^{T}S^{\mu }
{\rm p})(\pi _{\mu }+y^{T}S_{\mu }{\rm p})+V  
\label{eqn:3.13}
\end{eqnarray}
Since $S^\mu$ is antisymmetric, we may regard (\ref{eqn:3.13})
as a quantum Hamiltonian without ordering ambiguity in the
operator $y^T S^\mu {\rm p}$.  The Schr\"odinger equation is then
\begin{equation}
i\partial _{\tau }\psi ={\rm K}\psi =\left[ \frac{{\rm p}^{2}}{2m}+\frac{1}{
2mr_{0}^{2}}(\pi ^{\mu }+y^{T}S^{\mu }{\rm p})(\pi _{\mu }+y^{T}S_{\mu }{\rm 
p})+V\right] \psi ,  
\label{eqn:3.14}
\end{equation}
where we take as quantum operators
\begin{equation}
{\rm p}_\mu = -i \frac{\partial}{\partial y^\mu} \qquad \pi_\mu =
-i \frac{\partial}{\partial n^\mu} 
\label{eqn:3.15}
\end{equation}
We require that (\ref{eqn:3.14}) be locally gauge invariant
in the coordinate space $(n,y)$, that is, under transformations
of the form
\begin{equation}
\psi \longrightarrow e^{-ie\Theta(n,y)} \ \psi \ ;
\label{eqn:3.16}
\end{equation}
this can be accomplished through the minimal coupling
prescription
\begin{equation}
{\rm p}_\mu \longrightarrow {\rm p}_\mu -
e{\rm A}^{(n)}_\mu \qquad
\pi_\mu \longrightarrow \pi_\mu -  e\chi_\mu
\label{eqn:3.17}
\end{equation}
together with the requirement that under gauge transformation
\begin{equation}
{\rm A}^{(n)}_\mu \longrightarrow {\rm A}^{(n)}_\mu +
\frac{\partial}{\partial y^\mu}
\Theta \qquad
\chi_\mu \longrightarrow \chi_\mu +
(\frac{\partial}{\partial n^\mu}
+ y^T S_\mu \nabla_{\bf y}) \Theta.
\label{eqn:3.18}
\end{equation}
Note that ${\rm A}^{(n)}_\mu$ transforms under O(3,1) as an
induced (over O(2,1))
representation; it transforms as ${\rm p}_\mu$ under Lorentz
transformations (i.e., under the O(2,1) little group) and so,
since the Maxwell equations are Lorentz invariant,
it satisfies the Maxwell equation in the $y^\mu$
variables.  Under gauge transformation,
\begin{equation}
({\rm p} -e{\rm A}^{(n)\prime})e^{-ie\Theta}\psi =
e^{-ie\Theta}({\rm p}+e\nabla_{\bf y}
\Theta - e{\rm A}^{(n)\prime})\psi = e^{-ie\Theta}
({\rm p} - e{\rm A}^{(n)} )\psi
\label{eqn:3.19}
\end{equation}
and
\begin{eqnarray}
(\pi_\mu + y^T S_\mu {\rm p} -e\chi'_\mu )e^{-ie\Theta}\psi &=&
e^{-ie\Theta}
(\pi_\mu + y^T S_\mu {\rm p} + e \frac{\partial}{\partial n^\mu}
\Theta 
+ e y^T S_\mu \nabla_{\bf n} \Theta - e\chi'_\mu )\psi
\nonumber \\
&=& e^{-ie\Theta} (\pi_\mu + y^T S_\mu {\rm p} -e\chi_\mu )\psi ,
\label{eqn:3.20}
\end{eqnarray}
so that the gauge invariant form of (\ref{eqn:3.14}) is
\begin{equation}
i\partial _{\tau }\psi ={\rm K}\psi =\left[ \frac{1}{2m}({\rm p}-e{\rm A}
^{(n)})^{2}+\frac{1}{2mr_{0}^{2}}(\pi ^{\mu }+y^{T}S^{\mu }{\rm p}-e\chi
^{\mu })(\pi _{\mu }+y^{T}S_{\mu }{\rm p}-e\chi _{\mu })+V\right] \psi \ .
\label{eqn:3.21}
\end{equation}

Notice the operator 
\begin{equation}
D_\mu = \frac{\partial}{\partial n^\mu}  + y^T S_\mu \nabla_{\bf y}
= (\nabla_n)_\mu + y^T S_\mu \nabla_{\bf y}
\label{eqn:3.22}
\end{equation}
which appears in the second of (\ref{eqn:3.18}) and in (\ref{eqn:2.26}).  
For a function $f(n,y)$ defined  such that
its dependence on $n$ is only through ${\cal L}(n)^T y$ (which
is to say that $f$ is a function of $x$ alone, even as $n$
varies in $\tau$), we find that
\begin{equation}
\frac{\partial}{\partial y^\mu} f = \left.\frac{df}{d\xi^\alpha} 
\right|_{\xi={\cal L}(n)^T y} \frac{\partial}{\partial y^\mu}
({\cal L}_\beta^{\ \alpha} y^\beta) =
{\cal L}_\mu^{\ \alpha} \left.\frac{df}{d\xi^\alpha}
\right|_{\xi={\cal L}(n)^T y} 
\label{eqn:3.27}
\end{equation}
and
\begin{equation}
\frac{\partial}{\partial n^\mu} f = \left.\frac{df}{d\xi^\alpha}
\right|_{\xi={\cal L}(n)^T y} \frac{\partial}{\partial n^\mu}
({\cal L}_\beta^{\ \alpha} y^\beta)
\label{eqn:3.28}
\end{equation}
so that
\begin{eqnarray}
D_\mu f &=& \left(\frac{\partial}{\partial n^\mu} + y^T S_\mu
\nabla_{\bf y} \right) f 
\nonumber \\
&=& \left[\frac{\partial}{\partial n^\mu} + y_\beta
{\cal L}^\beta_{\ \gamma} (\frac{\partial}{\partial n^\mu}
{\cal L} ^{\alpha\gamma}) \frac{\partial}{\partial
y^\alpha}\right] f \nonumber \\
&=&  \left.\frac{df}{d\xi^\sigma} \right|_{\xi={\cal L}(n)^T y}
y^\beta
\left[ \frac{\partial}{\partial n^\mu}
{\cal L}_\beta^{\ \sigma}
+ {\cal L}_\beta^{\ \gamma}
( \frac{\partial}{\partial n^\mu} {\cal L} ^{\alpha}_{\ \gamma})
{\cal L}_\alpha^{\ \sigma}
\right] \nonumber \\
&=&  \left.\frac{df}{d\xi^\sigma} \right|_{\xi={\cal L}(n)^T y}
y^\beta
\left[
\frac{\partial}{\partial n^\mu}
{\cal L}_\beta^{\ \sigma} + {\cal L}_\beta^{\ \gamma}
({\cal L}^T)^\sigma_{\ \alpha} \frac{\partial}{\partial n^\mu}
{\cal L}^{\alpha}_{\ \gamma} 
\right] \nonumber \\
&=& \left.\frac{df}{d\xi^\sigma} \right|_{\xi={\cal L}(n)^T y}
y^\beta 
\left[ \frac{\partial}{\partial n^\mu} {\cal L}_\beta^{\ \sigma}
- {\cal L}_\beta^{\ \gamma} {\cal L} ^{\alpha}_{\ \gamma}
\frac{\partial}{\partial n^\mu} {\cal L}_\alpha^{\ \sigma}
\right] 
\nonumber \\ &\equiv& 0
\label{eqn:3.29}
\end{eqnarray}
where we have used (\ref{eqn:2.20}).
In fact, it follows from (\ref{eqn:2.25}) that
\begin{equation}
dx \cdot \nabla_{\bf x} + dn^\mu D_\mu = dy \cdot
\nabla_{\bf y} + dn \cdot \nabla_{\bf n}
\label{eqn:3.31}
\end{equation}
which shows that $\nabla_{\bf x}$ and $D_\mu$
generate the variations induced by $dx$ and $dn$, just
as $\nabla_{\bf y}$ and $\nabla_{\bf n}$ generate the 
variations induced by $dy$ and $dn$.
Thus, $D_\mu$ acts as a kind of covariant derivative which
vanishes on functions of $x$. 
In particular, $D_\mu$ vanishes on the eigenstates discussed in
\cite{I} and \cite{II}, in which case the Hamiltonian
(\ref{eqn:3.13}) reduces to the RMS Hamiltonian
discussed in \cite{I}.  

The classical Lagrangian associated with the locally gauge
invariant Hamiltonian (\ref{eqn:3.13}) is 
\begin{equation}
{\rm L} = \frac{1}{2} m \dot x^2 + \frac{1}{2}mr_{0}^{2}\dot{n}^{2}
+ e [ \dot x \cdot ({\cal L}^T {\rm A}^{(n)}) +
\dot n \cdot \chi ] - V(n,x).
\label{eqn:3.36}
\end{equation}
In order for L to be a Lorentz scalar, ${\cal L}^T
{\rm A}^{(n)}$ must transform under the full Lorentz
group O(3,1). 
Since ${\rm A}^{(n)}$ was introduced as a field which transforms
under the O(2,1) little group, we have that
\begin{equation}
{\rm A}^{(n)\prime} = D^{-1} (\Lambda , n) {\rm A}^{(n)} =
{\cal L} (\Lambda n) \; \Lambda \; {\cal L}^T (n) {\rm A}^{(n)}
\ .
\label{eqn:3.37}
\end{equation}
Operating on (\ref{eqn:3.37}) with ${\cal L}^T \; (\Lambda n)$
leads to
\begin{equation}
\Lambda \; \left[{\cal L}^T (n) {\rm A}^{(n)}\right] = 
{\cal L}^T \; (\Lambda n) {\rm A}^{(n)\prime} = 
\left[{\cal L}^T (n) {\rm A}^{(n)}\right]'
\label{vector}
\end{equation}
verifying that the combination ${\cal L}^T {\rm A}^{(n)}$
transforms as a four vector under $\Lambda$.

\section{Interaction With an External Field}
In (\ref{eqn:3.17}), we introduced the gauge compensation
fields, ${\rm A}^{(n)}_\mu$ and $\chi_\mu$, required to make the
Hamiltonian (\ref{eqn:3.13}) locally gauge invariant.  To avoid
introducing extra degrees of freedom,
we argue that just as $n$ and $y$ transform under
inequivalent representations of the Lorentz group ($y$ transforms
under the O(2,1) little group induced by the action of the full
O(3,1)), so ${\rm A}^{(n)}_\mu$
and $\chi_\mu$ should be seen as inequivalent representations of
the usual U(1) gauge group of electromagnetism.  In the full
spacelike region, a constant electromagnetic
field, $F^{\mu\nu}$, can be represented through the vector
potential
\begin{equation}
A^\mu (x) = -\frac{1}{2} F^{\mu\nu} x_\nu.
\label{eqn:4.5}
\end{equation}
We now restrict the support of $A^\mu$ to $x\in$ RMS($n$)
and express the vector potential as a vector oriented with
RMS(${\mathaccent'27n}$) by writing
\begin{equation}
{\rm A}^{(n)}_\mu (y)= {\cal L}_{\mu\nu} A^\nu ({\cal L}^T y)
=-\frac{1}{2} {\cal L}_{\mu\nu} F^\nu_{\ \sigma}
{\cal L}_\lambda^{\ \sigma} y^\lambda
=-\frac{1}{2} ({\cal L} F {\cal L}^T y)_\mu.
\label{eqn:4.6}
\end{equation}
For the field $\chi_\mu$, we choose (note
that $n$ undergoes Lorentz transform in the same way as $x$), 
\begin{equation}
\chi _{\mu }(n)=b^{2}\ A_{\mu }(n)=-\frac{b^{2}}{2}\ F_{\ \sigma }^{\nu }\
n^{\sigma }  
\label{eqn:4.7}
\end{equation}
(here $b$ is another length scale, required since $A_\mu (x)$ has
units of length$^{-1}$, so $F^\nu_\sigma$ must have units of
length$^{-2}$, but $\chi_\mu$ must be without units) and we use
(\ref{eqn:4.6}) and (\ref{eqn:4.7}) in the Schr\"odinger
equation (\ref{eqn:3.21}). 
\begin{eqnarray}
i\partial _{\tau }\psi &=&\left[ \frac{1}{2m}({\rm p}-e{\rm A}^{(n)})^{2}+
\frac{1}{2mr_{0}^{2}}(\pi ^{\mu }+y^{T}S^{\mu }{\rm p}-e\chi ^{\mu })(\pi
_{\mu }+y^{T}S_{\mu }{\rm p}-e\chi _{\mu })+V\right] \psi  \nonumber \\
&=&\left[ \frac{1}{2m}{\rm p}^{2}-\frac{e}{2m}({\rm p}\cdot {\rm A}^{(n)}+
{\rm A}^{(n)}\cdot {\rm p})+\frac{1}{2mr_{0}^{2}}(\pi ^{\mu }+y^{T}S^{\mu }
{\rm p})^{2}-\right.  \nonumber \\
&&\mbox{\quad }\left. \frac{e}{2mr_{0}^{2}}[(\pi ^{\mu }+y^{T}S^{\mu }{\rm p}
)\chi _{\mu }+\chi ^{\mu }(\pi _{\mu }+y^{T}S_{\mu }{\rm p})
]+V+o(e^{2})\right] \psi  \nonumber \\
&=&\left\{ \frac{1}{2m}{\rm p}^{2}+\frac{1}{2mr_{0}^{2}}(\pi ^{\mu
}+y^{T}S^{\mu }{\rm p})^{2}+V\right.  \nonumber \\
&&\mbox{\qquad }\left. -e\left[\frac{1}{m}{\rm A}^{(n)}\cdot {\rm p}+\frac{1}{
mr_{0}^{2}}\chi ^{\mu }(\pi _{\mu }+y^{T}S_{\mu }{\rm p})\right]+o(e^{2})\right\}
\psi  
\label{eqn:4.8}
\end{eqnarray}
where the first three terms of (\ref{eqn:4.8}) are the
unperturbed Hamiltonian ${\rm K}_0$.

The perturbation term to order $o(e)$, is
\begin{eqnarray}
-e\left[
\frac{1}{m} 
{\rm A}^{(n)}\cdot
{\rm p} \right.
&+&
\left. \frac{1}{mr_{0}^{2}}
\chi ^{\mu}
(
\pi _{\mu }
+y^{T}S_{\mu }{\rm p}
)
\right]  \nonumber \\
&=&-e\left[\frac{1}{m}{\rm A}^{(n)T}{\rm p}+\frac{1}{mr_{0}^{2}}
\left(\chi ^{T}\pi
+y^{T}(S\cdot \chi ){\rm p}\right)\right]  \nonumber \\
&=&-\frac{e}{2}[\frac{1}{m}({\cal L}F{\cal L}^{T}y)^{T}{\rm p}+\frac{b^{2}}{
mr_{0}^{2}}F_{\ \nu }^{\mu }n^{\nu }(\pi _{\mu }+y^{T}S_{\mu }{\rm p})] 
\nonumber \\
&=&\frac{e}{2m}[y^{T}{\cal L}F{\cal L}^{T}{\rm p}+\frac{mb^{2}}{mr_{0}^{2}}
n_{\nu }F^{\nu \mu }(\pi _{\mu }+y^{T}S_{\mu }{\rm p})].  
\label{eqn:4.9}
\end{eqnarray}
Expanding the electromagnetic field tensor on the basis of
four by four antisymmetric tensors given by the Lorentz
generators ${\cal M}^{\mu\nu}$,
\begin{equation}
F = \frac{1}{2} F_{\mu\nu} {\cal M}^{\mu\nu} \Longrightarrow
(F)^{\alpha\beta} = \frac{1}{2} F_{\mu\nu}
({\cal M}^{\mu\nu})^{\alpha\beta}
= \frac{1}{2} F_{\mu\nu} (g^{\mu\alpha}g^{\nu\beta} -
g^{\mu\beta}g^{\nu\alpha}) = F^{\alpha\beta}.
\label{eqn:4.10}
%\label{eqn:4.11}
\end{equation}
Using (\ref{eqn:4.10}) in (\ref{eqn:4.9}) we find that
the perturbation term to order $o(e)$ becomes
\begin{equation}
\frac{e}{4m}F_{\alpha \beta }[y^{T}{\cal L}{\cal M}^{\alpha \beta }{\cal L}
^{T}{\rm p}+\frac{b^{2}}{r_{0}^{2}}n_{\mu }({\cal M}^{\alpha \beta })^{\mu
\nu }(\pi _{\nu }+y^{T}S_{\nu }{\rm p})]  
\label{eqn:4.12}
\end{equation}
Taking $b = r_{0}$, then we may write the first
order perturbation (using (\ref{eqn:2.26})) as 
\begin{equation}
\frac{e}{4m}F_{\alpha \beta }[y^{T}{\cal L}{\cal M}^{\alpha \beta }{\cal L}
^{T}{\rm p}+n_{\mu }({\cal M}^{\alpha \beta })^{\mu \nu }(\pi _{\nu
}+y^{T}S_{\nu }{\rm p})]=\frac{e}{4m}F_{\alpha \beta }X^{\alpha \beta }.
\label{coupling}
\end{equation}

The interaction term in (\ref{coupling}) was used in \cite{zeeman} 
to obtain the Zeeman effect.  For the magnetic-like field with
$F^{\mu\nu}F_{\mu\nu} = 2({\bf B}^2 - {\bf E}^2) >0$, there
exists a frame for which the interaction is purely magnetic.  In
such a frame, the perturbation becomes
\begin{equation}
\frac{e}{4m} F_{\alpha\beta} X^{\alpha\beta} = 
\frac{e}{4m} F_{ij} X^{ij} = \frac{e}{4m} \epsilon_{ijk} B^k
X^{ij} = \frac{e}{2m} B^k \left[\frac{1}{2} \epsilon_{ijk}
X^{ij}\right]
= \frac{e}{2m} B^k h(\lambda_k)
\label{eqn:4.14}
\end{equation}

where $h(\lambda_k)$ are the three conserved generators of the
SU(2) rotation subgroup of SL(2,C) for the phase space
$\{(n,y);(\pi,{\rm p})\}$, that is, the angular momentum
operator for the eigenstates of the induced representation. 
Notice that in the matrix element for unperturbed eigenstates,
the second terms of (\ref{eqn:4.9}) vanishes, so the
relativistic Zeeman effect does not depend upon the values of
$r_0$ or $b$. 

In \cite{II}, the diagonal angular momentum operator is
$L_1(n) = h(\lambda_1)=-i\partial/\partial \gamma$, and
so if we take ${\bf B} =
B(1,0,0)$ then we find that
\begin{equation}
{\rm K}_0 \quad \longrightarrow \quad {\rm K} = {\rm K}_0 -
\frac{eB}{2m} h(\lambda_1)
\label{eqn:4.15}
\end{equation}
splits the mass levels of the bound states according to 
\begin{equation}
K_{\ell n} \quad \longrightarrow \quad K'_{\ell nq} = 
K_{\ell n} - \frac{eB}{2m} q
\label{eqn:4.16}
\end{equation}
In going from (\ref{eqn:4.15}) to (\ref{eqn:4.16}), we have used
the fact that the unperturbed Hamiltonian of (\ref{eqn:4.8})
reduces to the the unperturbed Hamiltonian of \cite{II}.
Equation (\ref{eqn:4.16}) further justifies the conclusion
reached in \cite{selrul} that $q$ is the magnetic quantum number.
As pointed out in \cite{II}, the quantum number $q$ belongs to a
representation in the double covering of the Lorentz group, which
takes on, in fact, half-integer value, and indicates even
multiplicity for the normal Zeeman splittings.  Moreover, the
manifest covariance of the formalism guarantees that the
splitting of the spectrum will be independent of the observer.

\section{The Stark Effect}

For the electric-like field with $F^{\mu\nu}F_{\mu\nu} <0$, we may
find a frame in which the interaction is purely electric, leading
to the covariant formulation of the Stark effect.  In this case,
we find from (\ref{coupling}) that the first order perturbation
is 
\begin{equation}
\frac{e}{4m}F_{\alpha \beta }X^{\alpha \beta }=\frac{e}{2m}
E^{j}\,ih_{n}\left( \lambda _{0j}\right)
\label{boost-couple}
\end{equation}
and the electric field couples to the boost generators, which
are off-diagonal, non-compact, and anti-Hermitian \cite{II}.  In
order to recover the usual Stark level splitting, we propose a
second contribution to the perturbation, given by the scalar
potential
\begin{equation}
V^{\prime }(x,n)=-e\left[ -\varepsilon ^{\mu }\,\left( x_{\mu
}+r_{0}n_{\mu }\right) \right] \ ,
\label{a_5}
\end{equation}
where $\varepsilon ^{\mu }$ is a constant four-vector.
Together, the perturbation is 
\begin{equation}
{\rm K}^{\prime }=\frac{e}{2m}E^{1}\,ih_{n}\left( \lambda _{01}\right)
+e\varepsilon ^{1}\,\left( x_{1}+r_{0}n_{1}\right) \ ,
\label{perturbation}
\end{equation}
where we have taken the fields along the $1$-axis.

We first consider separately the contribution from the usual
electric field; that is, we take $\varepsilon ^{1}=0$ in
(\ref{perturbation}).  The matrix elements for the boost
generators follow from directly their algebraic properties
\cite{II}, and so
\begin{eqnarray}
<n_{a^{\prime }}{\ell }^{\prime }n^{\prime }L^{\prime }q^{\prime}
c_{2}^{\prime }|&&\!\!\!\!\!\!\!\!\!ih_{n}\left( \lambda _{01}\right)
|n_{a}{\ell }nLqc_{2}> 
={\delta _{qq^{\prime }}\;\delta _{nn^{\prime }}\;\delta }_{{\ell }
^{\prime },{\ell }}{\delta }_{n_{a^{\prime },}n_{a}}\,\delta
(c_{2}-c_{2}^{\prime })\left.  \qquad\qquad\qquad\qquad \right.
\nonumber \\
\times
&&\!\!\!\!\!\!\!\!\!\left[ iC_{L}\sqrt{L^2-q^2}
\delta _{L^{\prime },L-1}
-iA_{L}q\delta_{L^{\prime},L} \right.
%\nonumber \\
%&&\qquad\qquad
- \left. iC_{L+1}\sqrt{(L+1)^{2}-q^{2}}
\delta _{L^{\prime},L+1}
\right] \,
\end{eqnarray}
where
\begin{eqnarray}
iC_{L} &=&-\frac{1}{L}\sqrt{\frac{\left( L^{2}-\hat{n}^{2}\right) \left(
L^{2}+c_{2}^{2}/\hat{n}^{2}\right) }{4L^{2}-1}} \\
iA_{L} &=&\frac{ic_{2}}{L(L+1)} \\
iC_{L+1} &=&-\frac{1}{L+1}\sqrt{\frac{\left( (L+1)^{2}-\hat{n}^{2}\right)
\left( (L+1)^{2}+c_{2}^{2}/\hat{n}^{2}\right) }{4(L+1)^{2}-1}} \
.
\end{eqnarray}
The contribution to the spectrum becomes
\begin{eqnarray}
<n_{a^{\prime }}{\ell }^{\prime }n^{\prime }L^{\prime }q^{\prime}
c_{2}^{\prime }|&&{\!\!\!\!\!\!\!\!\!} K^{\prime }|n_{a}{\ell }nLqc_{2}>
=
\frac{e}{2m}E\;\delta _{n_{a^{\prime }},n_{a}}\delta _{qq^{\prime
}}\;\delta _{nn^{\prime }}\;\delta _{{\ell }^{\prime },{\ell }}\,\delta
(c_{2}-c_{2}^{\prime }) 
\nonumber \\
&&{\!\!\!\!\!\!\!\!\!} 
\times\left[ -\frac{\sqrt{L^{2}-q^{2}}}{L}\sqrt{\frac{\left( L^{2}-\hat{n}
^{2}\right) \left( L^{2}+c_{2}^{2}/\hat{n}^{2}\right) }{4L^{2}-1}}\delta
_{L^{\prime },L-1}\right. 
-\frac{ic_{2}}{L(L+1)}\,q\,\delta _{L^{\prime },L} 
\nonumber \\
&&{\!\!\!\!\!\!\!\!\!} 
+\left. \frac{\sqrt{(L+1)^{2}-q^{2}}}{L+1}\sqrt{\frac{\left( (L+1)^{2}-
\hat{n}^{2}\right) \left( (L+1)^{2}+c_{2}^{2}/\hat{n}^{2}\right) }{
4(L+1)^{2}-1}}\delta _{L^{\prime },L+1}\right] \ .
\label{E-contrib}
\end{eqnarray}
We consider the contribution of this term to the ground state,
with the quantum numbers
\begin{equation}
n =0\quad \quad {\ell }={0}\quad n_{a}=0\quad {\ell +}n_{a}=0
\quad \quad L =1/2,3/2\quad q=\pm 1/2
\label{E-q-num}
\end{equation}
where we recall the even multiplicity for the relativistic
ground state.  Combining (\ref{E-contrib}) and (\ref{E-q-num}),
\begin{eqnarray}
<n_{a^{\prime }} {\ell }^{\prime }n^{\prime }L^{\prime }q^{\prime
}c_{2}^{\prime }|K^{\prime }|n_{a}{\ell }nLqc_{2}> 
&&{\!\!\!\!\!\!\!\!\!}
=\frac{eE}{2m}\delta _{n_{a^{\prime }},n_{a}}\delta _{qq^{\prime
}}\;\delta _{nn^{\prime }}\;\delta _{{\ell }^{\prime },{\ell }}\,\delta
(c_{2}-c_{2}^{\prime })
\nonumber \\
&&{\!\!\!\!\!\!\!\!\!} \times 
\left[ \;\left( -q\right) \;\left( \frac{4ic_{2}}{3}\delta _{L,\frac{1}{2}
}\delta _{L^{\prime },\frac{1}{2}}+\frac{4ic_{2}}{15}\delta _{L,\frac{3}{2}
}\delta _{L^{\prime },\frac{3}{2}}\right) \right. 
\nonumber \\
&& \left. +\frac{\sqrt{2}}{3}\sqrt{\frac{9}{4}+4c_{2}^{2}}\left( \delta _{L,
\frac{3}{2}}\delta _{L^{\prime },\frac{1}{2}}-\delta _{L,\frac{1}{2}}\delta
_{L^{\prime },\frac{3}{2}}\right) \right]
\end{eqnarray}
which after diagonalization, provides the following contribution
to the spectrum
\begin{equation}
\Delta K=\pm i\,\frac{eE}{2m}\left( \frac{6}{15}c_{2}\pm \sqrt{(\frac{4}{15}
c_{2})^{2}+\left( \frac{1}{4}+\frac{4}{9}c_{2}^{2}\right) }\right) 
\stackunder{c_{2}\rightarrow 0}{\longrightarrow } \ \pm i\,\frac{eE}{4m}
\label{E-cont}
\end{equation}
Since the contribution in (\ref{E-cont}) is pure imaginary, we
see that the usual electric field leads to the decay of the ground
state.

We now consider the scalar contribution to (\ref{perturbation});
that is we consider the case in which $E^1 = 0$.
The matrix elements for the operators $x^\mu$ and $n^\mu$ were
computed in \cite{selrul}, and the relevant results are 
\begin{eqnarray}
<n_{a^{\prime }}{\ell }^{\prime }n^{\prime }L^{\prime }q^{\prime
}c_{2}^{\prime }|x^{1}|n_{a}{\ell }nLqc_{2}> 
&=&<n_{a^{\prime }}{\ell }^{\prime }|\rho |n_{a}{\ell }>{{\frac{q}{L(L+1)}}
\delta _{qq^{\prime }}\;\delta _{nn^{\prime }}\;\delta _{L^{\prime }L}
}
\nonumber
\\ && \times {\sum_{i=\pm 1}E_{{\ell }n}^{(i)}\delta _{{\ell }^{\prime }\;{\ell }+i}\,}
\delta (c_{2}-c_{2}^{\prime })
\end{eqnarray}
where
\begin{equation}
{E_{{\ell }n}^{(i)}=}\left\{ 
\begin{array}{c}
\left( {\ell -}n+1\right) \sqrt{\frac{1}{2{\ell +1}}\frac{1}{2{\ell }
^{\prime }{+1}}\frac{\left( {\ell -}n\right) !}{\left( {\ell +}n\right) !}
\frac{\left( {\ell }^{\prime }{+}n\right) !}{\left( {\ell }^{\prime }{-}
n\right) !}},\qquad i=+1 \\ 
\\ 
\left( {\ell +}n\right) \sqrt{\frac{1}{2{\ell +1}}\frac{1}{2{\ell }^{\prime }
{+1}}\frac{\left( {\ell -}n\right) !}{\left( {\ell +}n\right) !}\frac{\left( 
{\ell }^{\prime }{+}n\right) !}{\left( {\ell }^{\prime }{-}n\right) !}}
,\qquad i=-1
\end{array}
\right.
\label{x_element}
\end{equation}
and
\begin{equation}
<n_{a^{\prime }}{\ell }^{\prime }n^{\prime }L^{\prime }q^{\prime
}c_{2}^{\prime }|n^{1}|n_{a}{\ell }nLqc_{2}>
= {\frac{q}{L(L+1)}}\delta _{qq^{\prime }}\;\delta _{nn^{\prime }}\;\delta
_{LL^{\prime }}\;\delta _{{\ell }{\ell }^{\prime }}\;\delta
_{n_{a}n_{a}^{\prime }}\;\delta (c_{2}-c_{2}^{\prime })\; .
\label{n_element}
\end{equation}
Collecting (\ref{a_5}), (\ref{x_element}), and (\ref{n_element}),
the perturbative contribution of the scalar term to the spectrum will be,
\begin{eqnarray}
<n_{a^{\prime }}{\ell }^{\prime }n^{\prime }L^{\prime }q^{\prime
}c_{2}^{\prime }|V^{\prime }|n_{a}{\ell }nLqc_{2}>  
&=&e\epsilon {\frac{q}{L(L+1)}}\delta _{qq^{\prime }}\;\delta _{nn^{\prime
}}\;\delta _{L^{\prime }L}\delta (c_{2}-c_{2}^{\prime })
\nonumber \\
&\times &\left[ <n_{a^{\prime }}{\ell }^{\prime }|\rho
|n_{a}{\ell }>\sum_{i=\pm
1}E_{{\ell }n}^{(i)}\delta _{{\ell }^{\prime }{\ell }+i}+r_0 \;
\delta _{{\ell }{
\ell }^{\prime }}\delta _{n_{a}n_{a^{\prime }}}\right] 
\label{scalar_pert}
\end{eqnarray}

Considering specifically the level splitting in the $2s-2p$
system, with the quantum numbers
\begin{equation}
n =0\quad L=1/2\quad q=\pm 1/2\quad
{\ell } ={0,1}\quad n_{a}=0,1\quad \ell +n_{a}=1
\label{eps-q-num}
\end{equation}
we combine (\ref{scalar_pert}) and (\ref{eps-q-num}) to find
\begin{eqnarray}
<n_{a^{\prime }}{\ell }^{\prime }n^{\prime }L^{\prime }q^{\prime
}c_{2}^{\prime }|V^{\prime }|n_{a}{\ell }nLqc_{2}>  
&=&e\epsilon \,{\rm sgn}(q)\delta _{qq^{\prime }}\;\delta _{nn^{\prime
}}\;\delta _{L^{\prime }L}\,\delta (c_{2}-c_{2}^{\prime })
\nonumber \\
&&\times\left[ \frac{2}{3}\;r_{0}\delta _{{\ell }{\ell }^{\prime }}+2a_{0}\left(
\delta _{{\ell }^{\prime },{\ell -1}}+\delta _{{\ell }^{\prime },{\ell +1}
}\right) \right] \ .
\end{eqnarray}
where $a_0 = \hbar^2 / (m e^2)$ is the Bohr radius, which enters
through the expectation value of $\rho$ with respect to the
radial wavefunctions.  After diagonalization, the contribution to
the spectrum is
\begin{equation}
\Delta K=\pm e\epsilon \,\left( \frac{2}{3}r_{0}\pm 2a_{0}\right) \ ,
\end{equation}
which may be
compared with the standard nonrelativistic result 
\begin{equation}
\Delta K_{\rm nonrelativistic}=\pm eE \,\left(3a_{0}\right) \ .
\end{equation}
The free parameter $r_{0}$ appears to be remain for comparison with
experiment.

\section{Interpretations}
The calculations in the previous section indicate that in first
order perturbation theory, the usual electric field has the
effect of causing the covariant bound state to decay, a
phenomenon known from the exact, non-perturbative treatment
of the Stark effect.  However, the observed shifting of the
spectral lines, understood semi-classically as the alignment of
the bound state's effective dipole moment in the external
electric field, is not reproduced from this contribution.  In order to
recover the usual Stark splitting, it was necessary to introduce
a scalar potential which depends linearly on the position
four-vector.  This scalar potential has a natural interpretation
in the pre-Maxwell electromagnetic theory, which we now present.

Consider the one particle Stueckelberg equation,
\begin{equation}
i\partial_\tau \psi(x,\tau) = 
\left[\frac{p_{\mu}p^{\mu}}{2M}
+ V(x) \right] \psi(x,\tau) \ .
\label{one_body}
\end{equation}
Saad, Horwitz, and Arshansky have argued \cite{saad} that the  
local gauge covariance of equation (\ref{one_body}) should  
include transformations which depend on $\tau$, as well as on  
the spacetime coordinates.  This requirement of full gauge  
covariance leads to a theory of five gauge compensation fields,
since gauge transformations are functions on the five dimensional
space $(x,\tau)$.  Under local gauge transformations of the form
\begin{equation}
\psi(x,\tau) \rightarrow e^{i e_{0} \Lambda (x,\tau)} \psi (x,\tau)
\label{eqn:5.11}
\end{equation}
the equation
\begin{equation}
-(i \partial_{\tau} -e_{0}a_{5})\psi(x,\tau)
=\frac{1}{2M}(p^{\mu}-e_{0}a^{\mu})(p_{\mu}-
e_{0}a_{\mu})\psi(x,\tau)
\label{eqn:5.12}
\end{equation}
is covariant, when the compensation fields transform as
\begin{equation}
a_{\mu}(x,\tau) \rightarrow
a_{\mu}(x,\tau)+\partial_{\mu}\Lambda(x,\tau) \qquad
a_{5}(x,\tau) \rightarrow a_{5}(x,\tau) + \partial_{\tau}
\Lambda(x,\tau) .
\label{eqn:5.14}
\end{equation}
The Schr\"odinger-like equation (\ref{eqn:5.12}) leads to the five
dimensional conserved current
\begin{equation}
\partial_{\mu}j^{\mu}+\partial_{\tau}j^{5}=0
\label{eqn:5.15}
\end{equation}
where
\begin{equation}
j^{5} = |\psi(x,\tau)|^{2} \qquad
j^{\mu} = \frac{-i }{2M}(\psi^{*}(\partial^{\mu}-i
e_{0}a^{\mu})\psi - \psi(\partial^{\mu}-i e_{0}a^{\mu})\psi^{*}).
\label{eqn:5.16}
\end{equation}
In analogy to nonrelativistic quantum mechanics the squared  
amplitude of the wave function may be interpreted as the  
probability of finding an event at $(\tau,x)$.  Equation  
(\ref{eqn:5.15}) may be written as  
$\partial_{\alpha}j^{\alpha}=0$, with $\alpha=0,1,2,3,5$.   

According to (\ref{eqn:5.12}), we can write the classical Hamiltonian as
\begin{equation}
K=\frac{1}{2M}(p^{\mu}-e_{0}a^{\mu})(p_{\mu}-
e_{0}a_{\mu})-e_{0}a_{5}
\label{eqn:h-cl}
\end{equation}
and using the Hamilton equations 
\begin{equation}
\frac{dx^{\mu}}{d\tau} =\frac{\partial K}{\partial p_{\mu}}
\qquad\qquad
\frac{dp^{\mu}}{d\tau} =-\frac{\partial K}{\partial x_{\mu}}
\label{eqn:5.8}
\end{equation}
we find
\begin{equation}
M \ \dot x^\mu = (p^{\mu}-e_{0}a^{\mu})
\label{eqn:p-cl}
\end{equation}
which enables us to write the classical Lagrangian,
\begin{eqnarray}
L &=& \dot x^\mu p_\mu - K
\nonumber \\
&=& \frac12 M \dot x^\mu \dot x_\mu + e_0 \dot x^\mu a_\mu
+ e_{0}a_{5}.
\label{eqn:l-cl}
\end{eqnarray}
We may find the Lorentz force
\cite{emlf} by applying the Euler-Lagrange equations to
(\ref{eqn:l-cl}), which in the notation
$\alpha,\beta = 0,1,2,3,5$,  is
\begin{equation}
M \: \ddot x^\mu = f^\mu_{\:\;\;\nu} \dot x^\nu + f^\mu_{\:\;\;5}
= f^\mu_{\:\;\;\alpha}(x,\tau) \, \dot x^\alpha . 
\label{eqn:5.18}
\end{equation}
where
\begin{equation}
f^{\mu\nu} = \partial^\mu a^\nu - \partial^\nu a^\mu
\qquad f^\mu_{\:\;\;5} = \partial^\mu a_5 - \partial_\tau a^\mu
\ .
\label{eqn:f-def}
\end{equation}
The four equations (\ref{eqn:5.18}) imply \cite{emlf} 
\begin{equation}
\frac{d}{d\tau} (\frac{1}{2} M \dot x^2) = M \dot x^\mu
\ddot x_\mu = \dot x^\mu (f_{\mu 5} + f_{\mu\nu} \dot
x^\nu ) = \dot x^\mu f_{\mu 5}
\label{eqn:5.n1}
\end{equation}
So, the conditions for the dynamical
conservation of $\dot x^2 = {\rm constant}$, are
\begin{equation}
f_{5\mu} = 0  \qquad {\rm and}  \qquad
\partial_\tau f^{\mu\nu}=0
\label{eqn:conditions}
\end{equation}
Thus, the mass-shell relation has the status, classically, of a
conservation law (a constant of motion conserved by Noether's
theorem for the $\tau$-translation symmetry) rather than a
constraint.

When we add as the dynamical term for the gauge field,
$(\lambda/4) f_{\alpha \beta}f^{\alpha \beta}$ 
where $\lambda$ is a dimensional constant,
the equations for the field are found to be
\begin{equation}
\partial_{\beta} f^{\alpha \beta}
=\frac{e_{0}}{\lambda}j^{\alpha}=ej^{\alpha}
\label{eqn:5.19}
\end{equation}
\begin{equation}
\epsilon^{\alpha \beta \gamma \delta
\epsilon}\partial_{\alpha}f_{\beta \gamma}=0
\label{eqn:ho}
\end{equation}
where $f_{\alpha\beta}=
\partial_{\alpha}a_{\beta}-\partial_{\beta}a_{\alpha}$, and
\begin{eqnarray}
j^\mu (\tau,y)&=& \dot x^\mu (\tau) \delta^4 \Bigl( y-x(\tau) \Bigr)
\label{eqn:c.45} \\
j^5 (\tau,y) &=& \rho (\tau,y)= \delta^4 \Bigl( y-x(\tau) \Bigr) \ .
\label{eqn:c.46}
\end{eqnarray}
We identify $e_0 / \lambda $ as the dimensionless Maxwell
charge (it follows from (\ref{eqn:5.22}) below that $e_0$ has
dimension of length).  The three vector form of the pre-Maxwell
equations are
\begin{eqnarray}
\nabla \cdot {\bf e} = ej^0 + \partial_\tau \varepsilon^0 &\qquad&   \nabla
\times {\bf e} + \partial_0 {\bf h} =0
\nonumber \\
\nabla \times {\bf h} - \partial_0 {\bf e} -\partial_\tau {\bf \varepsilon}
=e{\bf j}& \qquad&   \nabla \cdot {\bf h} =0
\nonumber \\
\nabla \cdot {\bf \varepsilon} = ej^4 - \partial_0 \varepsilon^0 &\qquad& \nabla
\times {\bf \varepsilon} - \sigma \partial_\tau {\bf h} =0
\nonumber \\
\nabla \varepsilon^0& =& -\sigma \partial_\tau {\bf e} -\partial_0 {\bf
\varepsilon}
\label{3-vec}
\end{eqnarray}
where
\begin{eqnarray}
e_i =f^{0i} &\qquad& h_i = {1\over 2} \epsilon_{ijk} f^{jk}
\nonumber \\
\varepsilon^i = f^{5i} &\qquad& \varepsilon^0 =f^{50}
\label{vec-defs}
\end{eqnarray}

Since the 4-vector part of the current in (\ref{eqn:5.16}) is not
conserved by itself, it may not be the source for the Maxwell field.
However, integration of (\ref{eqn:5.16}) over $\tau$,
with appropriate boundary conditions,
leads to $\partial_{\mu}J^{\mu}=0$, where
\begin{equation}
J^{\mu}(x)=\int_{-\infty}^{\infty} d\tau j^{\mu}(x,\tau)
\label{eqn:cat-j}
\end{equation}
so that we may identify $J^{\mu}$ as the source of the Maxwell
field.  Under appropriate boundary conditions, integration of
(\ref{eqn:5.19}) over $\tau$ implies
\begin{equation}
\partial_{\nu}F^{\mu \nu}=eJ^{\mu}
\qquad
\qquad
\epsilon^{\mu \nu \rho \lambda }\partial_{\mu}F_{\nu \rho}=0
\label{eqn:5.21}
\end{equation}
where
\begin{equation}
F^{\mu \nu}(x)=\int_{-\infty}^{\infty} d\tau f^{\mu \nu}(x,\tau)
\qquad
\qquad
A^{\mu}(x)=\int_{-\infty}^{\infty} d\tau a^{\mu}(x,\tau)
\label{eqn:5.22}
\end{equation}
so that $a^{\alpha}(x,\tau)$ has been called the pre-Maxwell field.  

In the pre-Maxwell theory, interactions take place between
events in spacetime rather than between worldlines.  
Each event, occurring at $\tau$, induces a current density in
spacetime which disperses for large $\tau$, and the continuity
equation (\ref{eqn:5.15}) states that these current densities
evolve as the event density $j^{5}$ progresses through
spacetime as a function of $\tau$.  As noted above, if
$j^{5}\rightarrow 0$ as $|\tau| \rightarrow \infty$ (pointwise
in spacetime), then the integral of
$j^{\mu}$ over $\tau$ may be identified with the Maxwell current.
This integration has been called
concatenation \cite{concat} and provides the link between the
event along a worldline and the notion of a particle, whose
support is the entire worldline.  Concatenation places the
electromagnetic field on the zero mass-shell.  The Maxwell
theory has the character of an {\it equilibrium limit} of the
microscopic pre-Maxwell theory.  

In consideration of the pre-Maxwell theory, the scalar
action-at-a-distance potential in the Horwitz-Piron quantum
theory, may be seen as an effective interaction resulting from
the scalar gauge potential $a_5$.  This effective interaction
follows from the concatenation process, by which microscopic
$\tau$-dependent evolution is averaged, according to
\begin{equation}
e_0 \ a_5(x,\tau) \qquad \stackunder{average}{\longrightarrow}
\qquad e_0 \ \ \left[\frac{1}{\lambda}
\int d\tau \ a_5(x,\tau) \right] = e A_5(x) = -V(x)
\label{conc-ave}
\end{equation}
so that the scalar potential plays the role of the Coulomb
potential in nonrelativistic mechanics.

If we consider a scalar potential of the form
\begin{equation}
V'(x) = -e \ A^5(x) = -e \ \varepsilon^\mu x_\mu
\end{equation}
with constant $\varepsilon^\mu$, then --- since $A^\mu(x)$ is
independent of $\tau$ --- the corresponding the field strength
tensor will be
\begin{equation}
F^{5\mu} = \partial^5 A^\mu - \partial^\mu A^5 = \varepsilon^\mu
\ .
\label{5-field}
\end{equation}
We see from (\ref{5-field}) that the choice of scalar potential
required to recover the Stark splitting from the covariant bound
state theory corresponds precisely to a constant external
four-vector electric field $F^{5\mu} =\varepsilon^\mu$, analogous
to the constant external three-vector electric field $F^{0j} =
E^j$ which causes the bound state to decay.  This interpretation
of the Stark effect calculation suggests that the parameterized
evolution theories of the Stueckelberg type require the
pre-Maxwell electromagnetic theory as a corollary, in order to
provide a complete description of known phenomenology.
% 
%%%%%%%%%%%%%%%%%%%%%%%%%% REFERENCES %%%%%%%%%%%%%%%%%%%%%%%%%%%%%
%

\end{document}